\pdfoutput=1

\documentclass[reprint,aps,prl,twocolumn,floatfix,superscriptaddress]{revtex4-1}
\usepackage{graphicx}
\usepackage{amssymb}
\usepackage{dcolumn}
\usepackage{bm}
\usepackage{amsmath,siunitx}
\usepackage{lineno}
\usepackage[english]{babel}
\usepackage[utf8]{inputenc}
\usepackage{xcolor}

\usepackage{blindtext}
\begin{document}

\title{A high-mobility hole bilayer in a germanium double quantum well}

\author{Alberto Tosato}
\affiliation{QuTech and Kavli Institute of Nanoscience, Delft University of Technology, PO Box 5046, 2600 GA Delft, The Netherlands}
\author{Beatrice M. Ferrari }
\affiliation{QuTech and Kavli Institute of Nanoscience, Delft University of Technology, PO Box 5046, 2600 GA Delft, The Netherlands}
\author{Amir Sammak}
\affiliation{QuTech and Kavli Institute of Nanoscience, Delft University of Technology, PO Box 5046, 2600 GA Delft, The Netherlands}
\affiliation {QuTech and TNO, Stieltjesweg 1, 2628 CK Delft, The Netherlands}
\author{Alexander R. Hamilton}
\affiliation{School of Physics, University of New South Wales, Sydney, New South Wales 2052, Australia}
\affiliation{ARC Centre of Excellence for Future Low-Energy Electronics Technologies, University of New South Wales, Sydney, New South Wales 2052, Australia}
\author{Menno Veldhorst}
\affiliation{QuTech and Kavli Institute of Nanoscience, Delft University of Technology, PO Box 5046, 2600 GA Delft, The Netherlands}
\author{Michele Virgilio}
\affiliation{Dipartimento di Fisica ``E. Fermi'', Universit\`a di Pisa,  Largo Pontecorvo 3, 56127 Pisa, Italy}
\author{Giordano Scappucci}
\email{g.scappucci@tudelft.nl}
\affiliation{QuTech and Kavli Institute of Nanoscience, Delft University of Technology, PO Box 5046, 2600 GA Delft, The Netherlands}

\date{\today}
\pacs{}

\begin{abstract}
We design, fabricate, and study a hole bilayer in a strained germanium double quantum well. Magnetotransport characterisation of double quantum well field-effect transistors as a function of gate voltage reveals the population of two hole channels  with a high combined mobility of $\SI{3.34e5}{cm^2.V^{-1}.s^{-1}}$ and a low percolation density of \SI{2.38e10}{cm^{-2}}. We resolve the individual population of the channels from the interference patterns of the Landau fan diagram. At a density of \SI{2.0e11}{cm^{-2}} the system is in resonance and we observe an anti-crossing of the first two bilayer subbands characterized by a symmetric-antisymmetric gap of $\sim\SI{0.69}{meV}$, in agreement with Schrödinger-Poisson simulations.

\end{abstract}

\maketitle

The development of high-quality undoped Ge/SiGe quantum wells~\cite{Sammak2019ShallowTechnology,Lodari2021LowGermanium} has established planar Ge as a front-runner material platform en route to a large-scale spin-qubit quantum processor~\cite{Scappucci2020TheRoute}. In only three years, key milestones have been demonstrated, such as stable and quiet quantum dots~\cite{Hendrickx2018Gate-controlledGermanium,Lodari2021LowGermanium} , single hole qubits~\cite{Hendrickx2020AQubit} with long relaxation times~\cite{Lawrie2020QuantumGermanium}, singlet-triplet qubits~\cite{Jirovec2021AGe}, fast two-qubit logic~\cite{Hendrickx2020FastGermanium}, universal operation on a $2\times2$ qubit array~\cite{Hendrickx2021AProcessor}, and simultaneous qubit driving at the fault-tolerant threshold~\cite{Lawrie2021SimultaneousThreshold}.
Exploiting the third dimension by integrating two (or more) quantum wells in the same heterostructure could provide extra degrees of freedom for designing an entire new class of quantum device architectures with tailored electronic properties. For example, quantum devices patterned in multiple layers may provide increased qubit connectivity for high performance quantum circuits. In these devices, the wavefunction of quantum confined holes may be shifted or delocalized in between quantum wells, providing a larger parameter space for effective mass, $g$-factor, and spin-orbit coupling tuning~\cite{Burkard2000SpinDots}. These are relevant parameters for advanced spin-qubit control~\cite{Scappucci2020TheRoute}. Furthermore, bilayers with high-mobility at low density may provide a suitable test bed for exploration of exotic phenomena such as exciton condensation~\cite{Eisenstein2004Bose-EinsteinSystems, Su2008HowFlow} and counterflow superconductivity in solid state devices at accessible temperatures~\cite{Rey2009ControlledSuperlattices, Conti2021ElectronholeHeterojunctions}.

\begin{figure}[!ht]
	\includegraphics[]{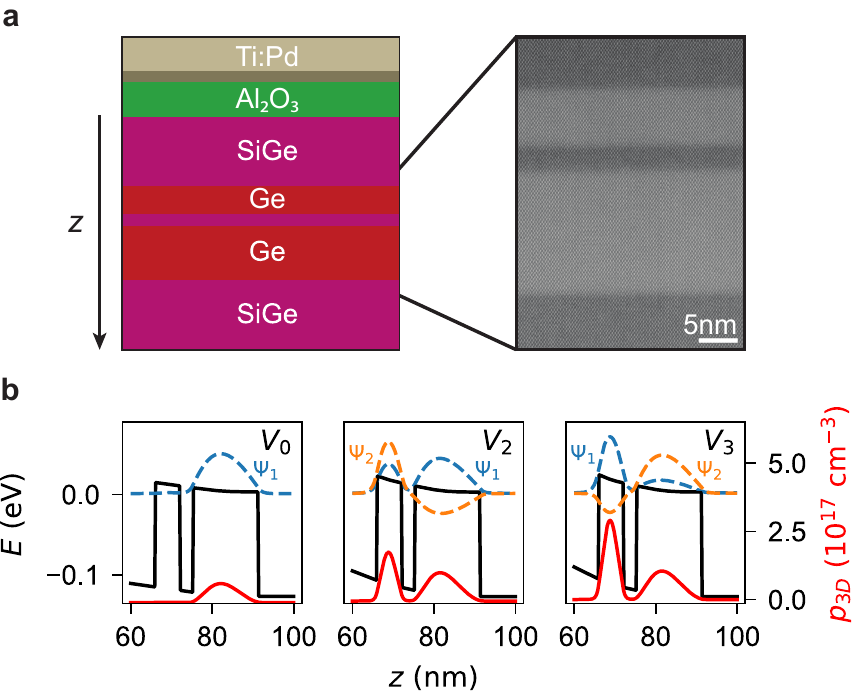}%
	\caption{\textbf{a} Schematics of the Ge/SiGe heterostructure with the gate stack and cross section of the double quantum well by transmission electron microscopy. \textbf{b} From left to right at increasing negative gate voltage ($V_0$, $V_2$, and $V_3$), each panel shows the heavy-holes band edge (black solid line), the wavefunction of the subbands above the Fermi energy (colored dashed lines), and the total density $p_{3\text{D}}=p_1 \lvert \Psi_1 \rvert ^2 + p_2 \lvert \Psi_2 \rvert ^2$ (red solid line) vs. bilayer depth ($z$). Here $\Psi_1$, $\Psi_2$ are the wavefunction amplitudes and $p_1$, $p_2$ the densities of the first and second subband. The Fermi energy is set as the reference energy at \SI{0}{eV}.}
\label{fig:heterostructure}
\end{figure}

Here we demonstrate hole bilayers in planar Ge double quantum wells with high mobility at low density, a first prerequisite for exploring any of these exciting avenues. Through careful design of the heterostructure and because of the low disorder in both quantum wells, we are able to study in detail the quantum transport properties of the system in the tunnel coupled regime and observe the signature of a symmetric-antisymmetric gap when we tune the density in the quantum wells to be the same.

Figure~\ref{fig:heterostructure}a shows a schematics of the heterostructure, along with a cross section of the double quantum well by transmission electron microscopy. The Ge/SiGe bilayer was grown on a 100~mm Si(001) substrate in a high-throughput reduced-pressure chemical vapor deposition reactor~\cite{Sammak2019ShallowTechnology}. The \SI{16}{nm}~bottom quantum well and the \SI{8}{nm}~top quantum well are separated by a thin \SI{3}{nm}~Si$_{0.2}$Ge$_{0.8}$ barrier. The bilayer is grown on a strain-relaxed Si$_{0.2}$Ge$_{0.8}$ buffer layer obtained by reverse grading and is separated from the gate-stack by a \SI{66}{nm}~Si$_{0.2}$Ge$_{0.8}$ barrier. The quantum wells are compressively strained, leading to states in the quantum wells with heavy hole (HH) symmetry  being lower energy than light holes (LH) states.

\begin{figure*}[!ht]
	\includegraphics[]{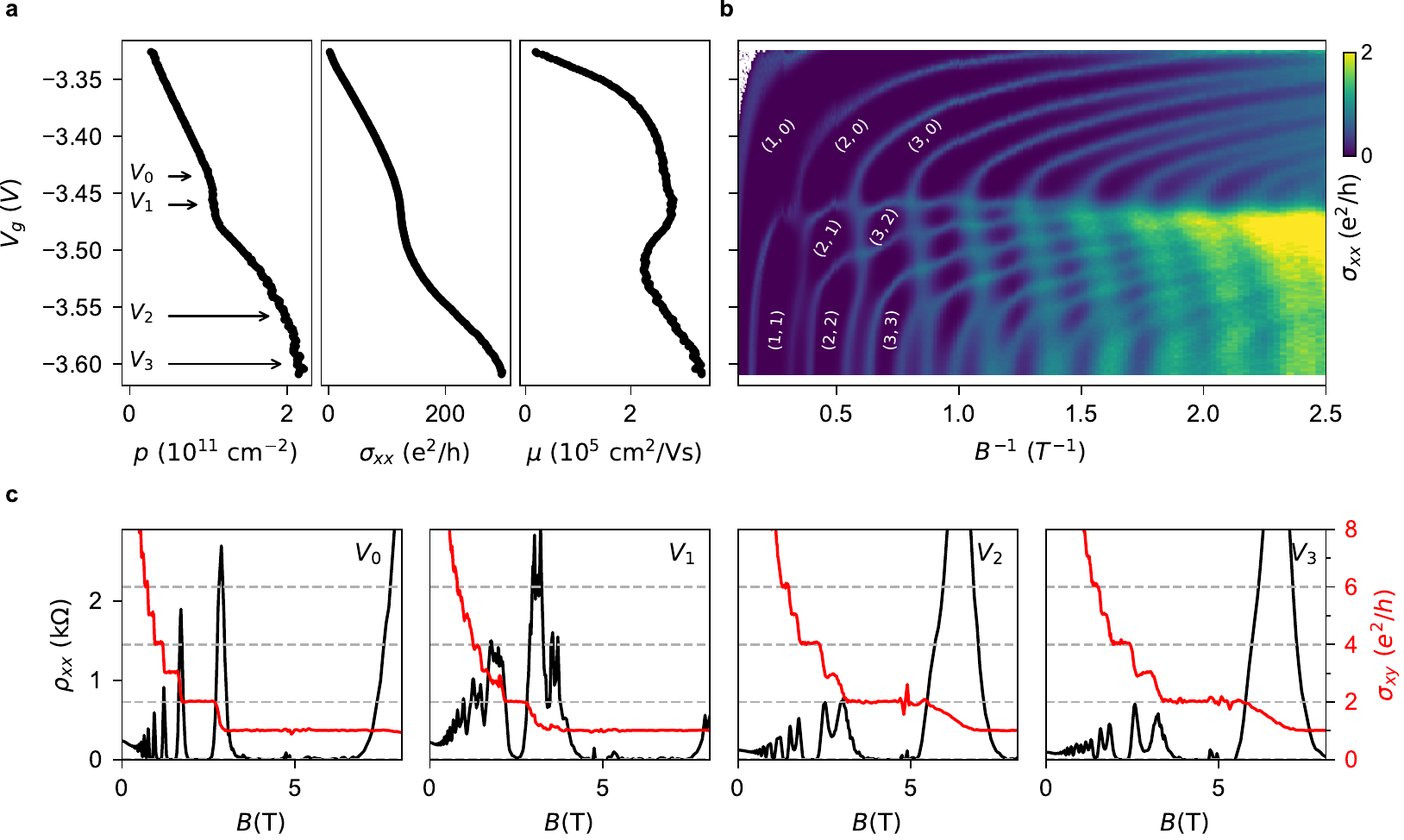}
	\caption{\textbf{a} From left to right: gate voltage ($V_g$) dependence of the hole bilayer Hall density $p$, conductivity $\sigma_{xx}$ at zero magnetic field , and mobility $\mu$. The bilayer behaviour at $V_g = V_0$, $V_1$, $V_2$, and $V_3$ is further described in the main text. \textbf{b} Colour map of the conductivity $\sigma_{xx}$ as a function of $V_g$ (same range as in \textbf{a}) and the inverse magnetic field $B^{-1}$. Dark regions correspond to filled Landau levels with vanishing $\sigma_{xx}$ and correspondingly quantized $\sigma_{xy}$. In brackets the filling factor $(\nu_{sbb1}, \nu_{sbb2})$ for the first and second subbands where $\nu=0$ indicates an empty subband. \textbf{c} From left to right: $\rho{xx}$ (black) and $\sigma_{xy}$ (red) as a function of magnetic field $B$ at $V_g = V_0$, $V_1$, $V_2$, and $V_3$. These line cuts are obtained from the magnetotransport measurement in \textbf{b}. $\sigma_{xy}$ shows plateaus at integer values of the total filling factor $(\nu_{sbb1}+ \nu_{sbb2})$  
	}
\label{fig:mgr}
\end{figure*}

The asymmetric design of the bilayer---the top quantum well being narrower than the bottom quantum well---allows both wells in the undoped heterostructure to be populated by applying a negative voltage to the top gate only~\cite{Laroche2015Magneto-transportHeterostructure}. We illustrate this capability in the three panels of Fig.~\ref{fig:heterostructure}b, which report the results of Schr\"{o}dinger-Poisson (SP) simulations of our bilayer for increasing negative gate voltages. Each panel shows the HH-band-edge, the total hole density and the wavefunction amplitude for the first $\Psi_1$ and second $\Psi_2$ subband of the bilayer system, as a function of the spatial coordinate z. At small gate voltages ($V=V_0$, left panel), only the first subband is populated and its wave function $\Psi_1$ is localized in the bottom well. At larger gate voltages occupation of the second subband becomes favourable with its wavefunction $\Psi_2$ initially localized in the top well. Then, the energy of the second subband increases with gate voltage until it anticrosses the energy level of the first subband at the resonance point. The central panel ($V=V_2$) shows the system at resonance. In this regime, the wavefunction of the first and second energy states are delocalized across both wells giving rise to the symmetric $\Psi_1$ and antisymmetric $\Psi_2$ states characteristic of a tunnel-coupled double quantum well system. The energy separation between the symmetric and antisymmetric states reaches its minimum ($\Delta_{SAS}$) at resonance. Upon further increasing the gate voltage ($V=V_3$ right panel), the wavefunction of the symmetric state shifts towards the top well and the total carrier density of the top quantum well increases while the one of the bottom quantum well remains unchanged.

We fabricated  Hall-bar shaped heterostructure field effect transistors (H-FETs) featuring platinum-germanosilicide ohmic contacts~\cite{Sammak2019ShallowTechnology} to the bilayer and performed measurements in a $^3$He dilution refrigerator with base temperature of \SI{50}{mK} and equipped with a \SI{12}{T} magnet. Standard voltage-bias four-probe lock-in technique at a frequency of \SI{17}{Hz} was used for magnetotransport characterization as described in ref.~\cite{Sammak2019ShallowTechnology}. We measure the longitudinal $\rho_{xx}$ and transverse $\rho_{xy}$ components of the resistivity tensor and via tensor inversion calculate the longitudinal $\sigma_{xx}$ and transverse $\sigma_{xy}$ conductivity.

The three panels of Fig.~\ref{fig:mgr}a show the zero-field longitudinal conductivity $\sigma_{xx}$, the bilayer density $p$ and the carrier mobility $\mu$ as a function of gate voltage $V_g$. At turn-on, only the bottom quantum well is populated. Mobility, density, and conductivity increase monotonically as the gate voltage is swept more negative up to a value $V_0 = \SI{-3.45}{V}$. In particular the Hall density increases linearly, consistent with a parallel-plate capacitor model of an H-FET with a single quantum well~\cite{Li2017EffectsHeterostructures}. We estimate a percolation density $p_p = \SI{2.38e10}{cm^{-2}}$  by fitting the conductivity to percolation theory  $\sigma_{xx} \sim (p-p_p)^{1.31}$ in the low density regime~\cite{Tracy2009ObservationMOSFET}. For $V_0 \leq V_g \leq V_1$ both the Hall density and conductivity deviate from the linear behaviour expected from SP simulations and flatten out. This observation signals that holes start populating the second subband; These holes are localized in the top quantum well, thereby screening the electric field at the bottom well. However, carriers in the top quantum well do not contribute to transport as their density is still below the percolation density. 
A further increase in negative gate voltage triggers transport in the top quantum well and for $V_1 \leq V_g \leq V_2$ we observe a transitory decrease in combined mobility due to inter-layer scattering~\cite{Stormer1982ObservationSystem}. For $V_g \geq V_2$ the combined mobility recovers its original monotonic increasing behaviour and saturates at $V_g = V_3$, reaching a maximum value $\mu = \SI{3.34e5}{cm^2.V^{-1}.s^{-1}}$ at bilayer density $p = \SI{2.21e11}{cm^{-2}}$. 

To elucidate the quantum transport properties of the bilayer we show in the right panel of Fig~\ref{fig:mgr}a a colour map of $\sigma_{xx}$ as a function of $V_g$ and the inverse magnetic field $B^{-1}$. Dark regions correspond to filled Landau levels (LLs) with vanishing $\sigma_{xx}$ and correspondingly quantized $\sigma_{xy}$. Bright lines correspond to sharp peaks in $\sigma_{xx}$ at half-filled LLs. Line cuts of $\rho_{xx}$ and $\sigma_{xy}$ at $V_g=V_0$, $V_1$, $V_2$, and $V_3$, are reported  in Fig~\ref{fig:mgr}b as a function of magnetic field. For $V_g \leq V_0$, the conductivity color map reveals a LL fan diagram typical of a single subband 2DHG, with Zeeman splitting resolved at $B^{-1}<2.5$~T$^{-1}$. This is highlighted in line cuts at $V=V_0$: $\rho_{xx}$ shows clean Shubnikov–de Haas (SdH) oscillations that vanish when $\sigma_{xy}$ develops flat conductance plateaus at integer multiples of $e^2/h$.

For $V_0\leq V_g\leq V_1$, the population of the second subband becomes favourable, charge starts accumulating in the top quantum well effectively screening the electric field in the bottom well. Although we observe the conductance peaks associated to the first subband saturating with gate voltage due to this screening effect, the fan diagram associated to the second subband appears only at $V_g=V_1$, in agreement with the observations in Fig.~\ref{fig:mgr}a, when the density in the second subband overcomes the percolation threshold and contributes to transport. Correspondingly, the line cuts for $\rho_{xx}$ and $\sigma_{xy}$ at $V_g=V_1$ show a complex pattern resulting from the parallel transport of two channels with different density and mobility.

\begin{figure}[!ht]
	\includegraphics[]{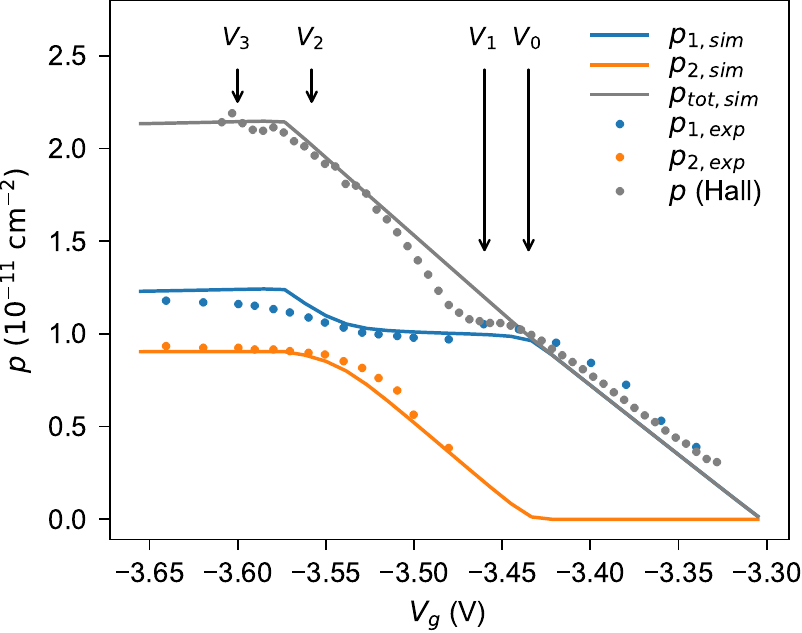}
	\caption{Comparison of the experimental and Schr\"{o}dinger-Poisson simulated densities in the bilayer system as a function of gate voltage ($V_g$). The blue and orange solid lines are the simulated density for the first and second subband, respectively. The gray solid line is the simulated total density in the bilayer. The blue and orange dotted lines are the density of the first and second subband, as computed from the position of the Shubnikov-de Haas conductance peaks in Fig.~\ref{fig:mgr}b. The dotted gray line is the bilayer Hall density reported in Fig.~\ref{fig:mgr}a.}
\label{fig:sim_exp}
\end{figure}

The hole bilayer reaches the resonance point at $V_g=V_2$, corresponding to a bilayer density of \SI{2.0e11}{ cm^{-2}}. At this point the wavefunction of the first and second subband are fully delocalized across both wells and we observe in the fan diagram the anticrossing of the $\sigma_{xx}$ peaks arising from the first subband with the corresponding peaks arising from the second subband. The anticrossing is better resolved at high magnetic field, where the LLs energy separation is large compared to the disorder-induced LLs broadening. The line cuts at resonance show a very distinct feature. A dip develops in $\rho_{xx}$ in correspondence of the  quantum Hall plateaus in $\sigma_{xy}$ at odd filling factors 3, 5 and 7. This observation is a clear signature of the anticrossing of symmetric-antisymmetric subbands at the resonance point in a tunnel coupled bilayer system~\cite{Hamilton1996FractionalSystems}. 
If the two quantum wells were not tunnel coupled, at resonance we would observe plateaus in $\sigma_{xy}$ only at even filling factors and a doubling of the $\rho_{xx}$ peaks height, arising from the measurement of two independent hole gases in parallel. 

Finally, as the negative gate voltage is further increased beyond resonance, the energy separation between the symmetric and antisymmetric gap increases; the symmetric subband rises in energy (and thus hole density) and the antisymmetric state remains relatively unchanged due to electric field screening effect. The line cuts at $V_g=V_3$ show a larger $\rho_{xx}$ peak separation in correspondence to larger odd Hall conductance plateaus. At $V_g \approx \SI{-3.61}{V}$ the bilayer system reaches its saturation as the triangular quantum well at the interface between the SiGe barrier and the dielectric starts populating, screening the electric field in both quantum wells from further increases in $V_g$.

A closer inspection of the fan diagram and line cuts reveals two interesting features of quantum transport in the bilayer.
The first observation is about the development of the $\sigma_{xx}$ peaks of the first subband in the color map. We note for  $V_1 \leq V_g \leq V_2$ that an increase in negative $V_g$ induces a shift of the peaks towards larger inverse magnetic field, implying that the charge in the first subband decreases as $V_g$ is swept further negative. We ascribe this behaviour to the negative compressibility of the second subband~\cite{Hamilton1996FractionalSystems}. When the charge density is still relatively low in the second subband, the first and second subbands are localized in the bottom and top quantum well, respectively. Making $V_g$ more negative must increase the total density. The negative compressibility of the low density gas hole in the top quantum well causes the density in the bottom quantum well to decrease immediately after the top well is populated, explaining why the $\sigma_{xx}$ peaks move to higher $B^{-1}$ (lower $B$) in the region $V_1 \leq V_g \leq V_2$~\cite{Eisenstein1992NegativeGases, Millard1995CompressibilitySystems}.

The second observation is about the emergence at resonance ($V_g=V_2$) of a quantum Hall state plateau at $\sigma_{xy}=1e^2/h$. This dissipationless state could be ascribed either to the full occupation of the symmetric state while the antisymmetric one remains empty, or it could arise from a fractional quantum hall state. At such high magnetic fields (above \SI{7}{T}), the intra-layer Coulomb energy $E_{intra} = e^2/\epsilon l_B$ (\SI{134}{meV} at \SI{8}{T}) is much bigger than the tunneling energy $\Delta_{SAS}$, as shown below, and the ratio of intra- to interlayer Coulomb energy $d/l_B =0.76$ in our bilayer is less than two, suggesting that at resonance, the quantum Hall state at total $\nu =1$ stems from spontaneous interlayer phase coherence~\cite{Yang1996SpontaneousSolitons}. Here, $d$ is the distance between the centres of the two wells.

The accurate measurement of the LL fan diagram allows to compare in Fig.~\ref{fig:sim_exp} the experimental density of the first and second subband to simulations~\footnote{We performed iterative self-consistent SP simulations of this Ge/SiGe bilayer system. We set in the model the thicknesses of SiGe spacer above the bilayer and Al$_2$O$_3$, to match the saturation density and capacitance respectively, of the measured device.}. We calculate the experimental subband density $p_1$ and $p_2$ using the quantum Hall relationship $p=\nu B e/h$ and tracking in Fig~\ref{fig:mgr}b the $B$ position of the $\sigma_{xx}$ peaks corresponding to the half filled LL with $\nu=1.5$~\footnote{Using the peaks of the half filled LL with $\nu=2.5$ yields a very similar density.}. 
We observe that the simulated subband population ($p_{1,sim}$, $p_{2,sim}$) matches well the experimental densities ($p_{1,exp}$, $p_{2,exp}$) in the ranges $V_g \le V_0$ and $V_g \ge V_2$. SP calculations only include the Hartree contribution to many body effects, and so do not reproduce the negative compressibility effects observed in the range $V_1 \leq V_g \leq V_2$; we therefore expect a discrepancy between simulations and experimental data in this region. Furthermore, in the range between $V_0$ and $V_1$, we ascribe the deviation of the measured Hall density $p$ from the simulated bilayer density $p_{tot,sim}$, to the population of the second subband when its density is still below the percolation threshold and effectively does not contribute to transport.

Around resonance ($V_g=V_2$) we observe the avoided-crossing of the simulated and experimental densities of the first and second subband. From the subband population difference at resonance $\Delta p_{SAS}=p_1-p_2$ we calculate the experimental energy gap between the symmetric and antisymmetric subbands by dividing $\Delta p_{SAS}$ by the spin-resolved 2D density of states $m^*/2\pi \hbar^2$. We obtain $\Delta_{SAS} \sim \SI{0.69}{meV}$, which is in agreement with simulations. 

In summary, we have designed, fabricated, and measured a hole bilayer in strained Ge double quantum wells. The bilayer has high-mobility, low percolation density, and a large symmetric-antisymmetric energy gap at resonance in agreement with simulations. Taken together, these results open up a plethora of exciting new possibilities for the Ge quantum information route, ranging from Ge quantum devices and circuits with increased connectivity to the exploration in this new platform of rich physics associated with quantum Hall effects in bilayers.

Data sets supporting the findings of this study are available at https://doi.org/10.4121/17209091

\bibliography{refs.bib}

\end{document}